\documentclass[12pt]{article}
\usepackage{graphicx}
\usepackage{epsfig}
\usepackage{epstopdf}
\DeclareGraphicsExtensions{.pdf,.eps,.png,.jpg,.mps}

\usepackage{cite}

\def\lsim{\mathrel{\rlap{\lower3pt\hbox{\hskip0pt$\sim$}}
     \raise1pt\hbox{$<$}}}         %less than or approx. symbol
\def\gsim{\mathrel{\rlap{\lower4pt\hbox{\hskip1pt$\sim$}}
     \raise1pt\hbox{$>$}}}         %greater than or approx. symbol

\usepackage{amsmath}
\usepackage{amsfonts}

\begin{document}
\begin{titlepage}

\centerline{\Large \bf Dead Alphas as Risk Factors}
\medskip

\centerline{Zura Kakushadze$^\S$$^\dag$\footnote{\, Zura Kakushadze, Ph.D., is the President of Quantigic$^\circledR$ Solutions LLC,
and a Full Professor at Free University of Tbilisi. Email: zura@quantigic.com} and Willie Yu$^\sharp$\footnote{\, Willie Yu, Ph.D., is a Research Fellow at Duke-NUS Medical School. Email: willie.yu@duke-nus.edu.sg}}
\bigskip

\centerline{\em $^\S$ Quantigic$^\circledR$ Solutions LLC}
\centerline{\em 1127 High Ridge Road \#135, Stamford, CT 06905\,\,\footnote{\, DISCLAIMER: This address is used by the corresponding author for no
purpose other than to indicate his professional affiliation as is customary in
publications. In particular, the contents of this paper
are not intended as an investment, legal, tax or any other such advice,
and in no way represent views of Quantigic$^\circledR$ Solutions LLC,
the website \underline{www.quantigic.com} or any of their other affiliates.
}}
\centerline{\em $^\dag$ Free University of Tbilisi, Business School \& School of Physics}
\centerline{\em 240, David Agmashenebeli Alley, Tbilisi, 0159, Georgia}
\centerline{\em $^\sharp$ Centre for Computational Biology, Duke-NUS Medical School}
\centerline{\em 8 College Road, Singapore 169857}
\medskip
\centerline{(July 31, 2017)}

\bigskip
\medskip

\begin{abstract}
{}We give an explicit algorithm and source code for extracting equity risk factors from dead (a.k.a. ``flatlined" or ``hockey-stick") alphas and using them to improve performance characteristics of good (tradable) alphas. In a nutshell, we use dead alphas to extract directions in the space of stock returns along which there is no money to be made (and/or those bets are too volatile). In practice the number of dead alphas can be large compared with the number of underlying stocks and care is required in identifying the aforesaid directions.
\end{abstract}
\medskip
\end{titlepage}

\newpage
\section{Introduction and Summary}

{}In the olden days quantitative trading (e.g., statistical arbitrage) strategies typically were based on some kind of an underlying idea that had to ``make sense". Because of this, the number of trading strategies back then (excluding evident variations) was relatively limited. But then hardware got really cheap and quant traders turned to data mining\footnote{\, See, e.g, \cite{Tulchinsky}.} which resulted in exponential proliferation of alphas\footnote{\, Here ``alpha" -- following the common trader lingo -- generally means any reasonable ``expected return" that one may wish to trade on and is not necessarily the same as the ``academic" alpha.  In practice, often the detailed information about how alphas are constructed may not even be available, e.g., the only data available could be the position data, so ``alpha" then is a set of instructions to achieve certain stock (or some other instrument) holdings by some times $t_1,t_2,\dots$} in modern quant trading,\footnote{\, See, e.g., \cite{4K}, \cite{101}.} and these ever-ubiquitous alphas are ever fainter and more ephemeral.

{}Alphas die -- that is, ``flatline" or become so faint that they are no longer tradable (a.k.a. ``hockey-stick" alphas) -- and are replaced with new data-mined alphas all the time, some or all of which in time will likely suffer the same fate. Typically, one simply tosses these dead alphas. In this note we ask the following question:

{}{\em Can we use the information encoded in dead alphas to improve the performance of good (tradable) alphas?} Put differently, can we learn something from dead alphas that we could possibly use to make good alphas more profitable and/or less volatile, etc.? As we discuss in the following, the answer to this question is affirmative. Thus, dead alphas define directions in the space of stock returns along which there is no money to be made (and/or those bets are too volatile). And this is valuable information. We can use it to make our good alphas even better. How? The idea here is simple. We can factor out these directions from good alphas, i.e., tweak them such that they do not make bets in these unproductive directions even partially.

{}In terms of a practical implementation, to identify the aforesaid directions in the space of stock returns associated with dead alphas we use the underlying stock\footnote{\, Our approach is agnostic to the nature of the underlying tradables, which need not be stocks.} position data, which is known for both dead and good alphas. What we end up extracting from this data is essentially a {\em risk factor loadings matrix} using which we can neutralize (via, e.g., a regression) stock positions of good alphas or build a multifactor risk model for optimizing our stock portfolio. This stock portfolio is obtained i) either via combining good alphas with some weights (see, e.g., \cite{Billion}), or ii) by calculating the expected returns for stocks using the expected returns for good alphas as in \cite{Decode} and directly trading a portfolio of stocks based on these expected returns (without combining alphas).

{}In Section \ref{sec.2} we give an explicit algorithm for computing the directions associated with dead alphas and factoring them out of good (tradable) alphas. We give the source code for this algorithm in Appendix \ref{app.A}.\footnote{\, The source code given in Appendix \ref{app.A} hereof is not written to be ``fancy" or optimized for speed or in any other way. Its sole purpose is to illustrate the algorithms described in the main text in a simple-to-understand fashion. Some important legalese is relegated to Appendix \ref{app.B}.} We briefly conclude in Section \ref{sec.3}.

\section{(Dead) Alphas}\label{sec.2}

{}Our discussion below a priori is oblivious to the underlying instruments. In practice, the most important application is in the case of equity portfolios comprised of a large number of stocks (e.g, 2,000+ most liquid U.S. stocks). So, for definiteness let us focus on alphas that trade (largely) overlapping portfolios of US stocks. Let the number of all stocks traded be $M$, and let the number of alphas be $N \gg M$. Let the realized returns for alphas be $\rho_{is}$ ($i=1,\dots,N$), and the realized returns for stocks be $R_{As}$ ($A=1,\dots,M$), where the index $s=1,\dots,T$ labels trading days (for definiteness, let $s=1$ correspond to the most recent date). Then we have
\begin{equation}\label{ex-post}
 \rho_{is} = \sum_{A=1}^M P_{iAs}~R_{As}
\end{equation}
Here on each day labeled by $s$ for each alpha labeled by $i$ the quantities $P_{iAs}$ are nothing but the properly normalized stock positions in the corresponding alpha portfolio. The normalization condition is given by
\begin{equation}\label{ov.norm}
 \sum_{A=1}^M |P_{iAs}| = 1
\end{equation}
There may be additional {\em linear constraints} on stock positions stemming from all alphas being subject to the same risk management restrictions. E.g., suppose all alphas are dollar-neutral. Then we have:
\begin{equation}\label{dollar.neutral}
 \sum_{A=1}^M P_{iAs} \equiv 0
\end{equation}
More generally, we can have $p$ linear constraints (typically, $p \ll M$, so we assume that this holds)
\begin{equation}\label{lin.con}
 \sum_{A=1}^M P_{iAs}~Q_{A\alpha} \equiv 0,~~~\alpha = 1,\dots,p
\end{equation}
Without loss of generality we can assume that: i) the $p$ columns of the matrix $Q_{A\alpha}$ are linearly independent;\footnote{\, The constraints (\ref{lin.con}) are invariant under $SO(p)$ rotations $Q\rightarrow Q~U$, where $U_{\alpha\beta}$ is an orthogonal matrix: $U U^T = U^T U = 1$.} and ii) the constraints (\ref{lin.con}) do not imply that $P_{iAs} \equiv 0$ for any given value of $A$ (or else none of the alphas trade the stock labeled by $A$ and we can simply drop it out of the universe). Furthermore, to keep it simple, we will assume that if there is a subset of $N_1$ alphas ($N_1 < N$) for which -- but not for all alphas -- we have additional linear constraints other than (\ref{lin.con}), then $N_1 \ll N$.

{}Not all alphas are viable for trading. Thus, suppose we take some time horizon, e.g., a period comprised of $d$ trading days. Then we can define {\em expected} returns for alphas, call them $\eta_{is}$, via moving averages based on prior $d$ days' worth of realized returns:\footnote{\, We emphasize that this is only an example and there are other ways of constructing $\eta_{is}$.}
\begin{equation}\label{eta}
 \eta_{is} = {1\over d} \sum_{s^\prime = s + 1}^{s + d} \rho_{is^\prime}
\end{equation}
Ideally, we want these expected returns to be i) positive, ii) not too low, and iii) not too volatile. We can quantify ii) and iii) as follows. E.g., we can demand that $\eta_{is} \geq \eta_{min}$, where $\eta_{min}$ is some lower bound on the expected returns below which alphas are no longer worth trading. Also, e.g., we can demand that the Sharpe ratios $\eta_{is} / \sigma_{is} \geq S_{min}$, where $S_{min}$ is some lower bound below which alphas are too volatile. And so on. Here ($\mbox{Var}(\cdot, d)$ is a $d$-day moving {\em serial} variance):
\begin{eqnarray}\label{var.d}
 &&\sigma_{is}^2 = \mbox{Var}(\eta_{is}, d) = {1\over{d-1}}\sum_{s^\prime = s+1}^{s+d} (\eta_{is^\prime} - {\overline \eta}_{is})^2\\
 &&{\overline \eta}_{is} = {1\over d}\sum_{s^\prime = s+1}^{s+d} \eta_{is^\prime}
\end{eqnarray}
So, we can define {\em dead} alphas as those for which, e.g., $\eta_{is} < \eta_{dead}$ and $\eta_{is} / \sigma_{is} < S_{dead}$, where $\eta_{dead}$ and $S_{dead}$ can -- but need not -- be the same as $\eta_{min}$ and $S_{min}$. Then, similarly, good (tradable) alphas are those for which $\eta_{is} \geq \eta_{min}$ and $\eta_{is} / \sigma_{is} \geq S_{min}$.

\subsection{Dead Alphas as Risk Factors}

{}Let us denote the subset of dead alphas as $J^\prime$, i.e., for $i\in J^\prime$ the corresponding alpha is dead. We are not going to trade these alphas. However, simply tossing them would be wasteful too. We can use them to improve good alphas we are trading -- let the subset of these alphas be $J$. We then have $N_{dead} + N_{good} \leq N$, where $N_{dead} = |J^\prime|$ is the number of dead alphas, while $N_{good} = |J|$ is the number of good alphas.\footnote{\, We are assuming that $\eta_{dead} \leq \eta_{min}$ and $S_{dead} \leq S_{min}$. Also, in principle, some dead alphas may start performing well again down the road, so one should bear this in mind.}

{}Dead alphas define directions in the space of stock returns along which there is no money to be made (and/or those bets are too volatile). This is valuable information. We can use it to make our good alphas even better. How? The idea here is simple. We can factor out these directions from good alphas. I.e., we can tweak good alphas such that they do not make bets in these unproductive directions even partially.

{}However, first we must identify these directions. The issue is that, precisely because of proliferation and ephemeral nature of alphas, typically, the number of dead alphas $N_{dead}$ is large: $N_{dead} \gg M$. It is clear that we cannot factor out more directions than we have stocks. So, then, how can we define these directions?

{}Here is a simple solution. Let us define the $M\times M$ matrix (for each value of $s$):
\begin{equation}\label{X}
 X_{ABs} = \sum_{i \in J^\prime} P_{iAs}~P_{iBs}
\end{equation}
In the presence of the linear constraints (\ref{lin.con}), this matrix is singular; otherwise, if $N_{dead} \gg M$, generically it is nonsingular. (In the remainder of this paragraph, to simplify notations, we suppress the index $s$.) We can compute the first $K$ principal components $V^{(a)}_A$, $a = 1,\dots,K$, of $X_{AB}$ (with the corresponding eigenvalues in the descending order, $\lambda^{(1)} > \lambda^{(2)} > \dots > \lambda^{(K)}$), where we have to determine what to take for $K$. A simple method is to take $K$ to be the {\em effective rank} (or eRank)\footnote{\, Here $\mbox{eRank}(Z)$ of a symmetric semi-positive-definite matrix (which suffices for our purposes here) $Z$ is defined as $\mbox{eRank}(Z) = \exp(H)$, where $H = -\sum_{a=1}^L p_a\ln(p_a)$ has the meaning of the (Shannon a.k.a. spectral) entropy \cite{Campbell60}, \cite{YGH}, $p_a = \lambda^{(a)} / \sum_{b=1}^L \lambda^{(b)}$, and $\lambda^{(a)}$ are the $L$ {\em positive} eigenvalues of $Z$.} of the matrix $X_{AB}$ \cite{RV}. More precisely, since eRank is fractional, we can take, e.g., $K = \mbox{Round}(\mbox{eRank}(X_{AB}))$ or $K = \mbox{floor}(\mbox{eRank}(X_{AB})) = \lfloor\mbox{eRank}(X_{AB})\rfloor$.

{}So, restoring the index $s$, we have $K_s$ principal components $V^{(a)}_{As}$, $a = 1,\dots,K_s$, which define the directions along which we wish to avoid taking positions in our trading portfolio. A simple way to achieve this is to include the $K_s$ vectors $V^{(a)}_{As}$ as columns in the linear constraints matrix $Q_{A\alpha}$ in (\ref{lin.con}) for {\em good alphas} labeled by $i\in J$. I.e., we simply neutralize good alphas w.r.t. $V^{(a)}_{As}$. A simple way to do this is, for each value of $i \in J$ and each value of $s$, to run a linear regression without the intercept of $P_{iAs}$ over $V^{(a)}_{As}$ and (up to overall normalization factors $\gamma_{is}$ -- see below) to take the residuals of this regression as the new desired holdings, call them ${\widetilde P}_{iAs}$:
\begin{equation}
 {\widetilde P}_{iAs} = \gamma_{is}\left[P_{iAs} - \sum_{a = 1}^{K_s} \sum_{B=1}^M V^{(a)}_{As}~V^{(a)}_{Bs}~P_{iBs}\right]
\end{equation}
Note that we have
\begin{equation}
 \sum_{A=1}^M {\widetilde P}_{iAs}~V^{(a)}_{As} \equiv 0
\end{equation}
which is due to the orthogonality of the principal components:
\begin{equation}
 \sum_{A=1}^M V^{(a)}_{As}~V^{(b)}_{As} = \delta_{ab}
\end{equation}
Here $\delta_{ab}$ is the Kronecker delta. The overall normalization factors $\gamma_{is}$ are fixed via
\begin{equation}
 \sum_{A=1}^M |{\widetilde P}_{iAs}| = 1
\end{equation}
which are the same normalization requirements as (\ref{ov.norm}) for the original positions $P_{iAs}$.

\subsection{A Tweak}

{}Above we compute the matrix $X_{ABs}$ via (\ref{X}), i.e., using the positions $P_{iAs}$ for dead alphas for one trading day (which is the trading day on which we are going to achieve the desired holdings ${\widetilde P}_{iAs}$ for good alphas). If we rebuild the risk model (see below) daily, then this is fine. However, this could also introduce additional twitch (turnover) into our desired trades. To mitigate this, we can average such matrices over some number trading days $d$ (which can -- but need not -- be the same as $d$ in (\ref{eta}) or (\ref{var.d})). I.e., instead of (\ref{X}) we take
\begin{equation}\label{X1}
 X_{ABs} = {1\over d} \sum_{s^\prime = s}^{s+d-1} \sum_{i \in J^\prime} P_{iAs}~P_{iBs}
\end{equation}
The rest goes exactly as above. Let us note that while (\ref{X}) and (\ref{X1}) at first look ``in-sample" as they are based on the data for the date $s$ for which we are computing ${\widetilde P}_{iAs}$, there is nothing improper about this: $P_{iAs}$ for both dead and good alphas are known in advance of the commencement of trading, that is, they are ``previsible".

\section{Concluding Remarks}\label{sec.3}

{}Instead of outright neutralizing the desired holdings w.r.t. $V^{(a)}_{As}$ as above (which amounts to regressing $P_{iAs}$ over $V^{(a)}_{As}$ and taking the residuals, up to overall normalization factors), we can take a different route. We can take good alphas labeled by $i\in J$ and i) either combine them with some weights (see, e.g., \cite{Billion}) and trade the resultant portfolio, or ii) calculate the expected returns for stocks using the expected returns for alphas as in \cite{Decode} and directly trade a portfolio of stocks based on these expected returns (without combining alphas). In the case i) above we get a stock portfolio which can be further optimized, e.g., by maximizing the Sharpe ratio \cite{Sharpe94} or via mean-variance optimization \cite{Markowitz}. For this (as well as in the case ii) above) we need to construct a risk model covariance matrix -- call it $\Phi_{AB}$ -- for our stock portfolio:\footnote{\, See, e.g., \cite{GK} for a general discussion. For an explicit open-source implementation of a general multifactor risk model for equities, see \cite{HetPlus}.}
\begin{equation}
 \Phi_{AB} = \xi_A^2~\delta_{AB} + \sum_{\mu,\nu=1}^F \Omega_{A\mu}~\phi_{\mu\nu}~\Omega_{B\nu}
\end{equation}
where $\xi_A$ is the specific (idiosyncratic) risk, $\Omega_{A\mu}$ is the risk factor loadings matrix, and $\phi_{\mu\nu}$ is the factor covariance matrix for $F$ risk factors. So, the idea here is that we can include the $K_s$ vectors $V^{(a)}_{As}$ as columns in $\Omega_{A\mu}$. Then, once we optimize using the so-constructed $\Phi_{AB}$, the resulting stock portfolio is only approximately neutral w.r.t. $V^{(a)}_{As}$. Exact neutrality is attained in the $\xi_A\rightarrow 0$ limit \cite{MeanRev}.

{}Whether we neutralize or optimize w.r.t. to the directions defined by dead alphas, the upshot is that dead alphas are actually useful. While they no longer make money, if used cerebrally, they help make more money and/or reduce portfolio volatility by eliminating/minimizing taking positions along these direction when trading good alphas. Finally, we present no backtests here as position data is highly proprietary.

\appendix

\section{R Source Code}\label{app.A}

{}In this appendix we give the R source code\footnote{\, The R Project for Statistical Computing, www.r-project.org.} for computing the directions associated with dead alphas and factoring them out of good (tradable) alphas. The code below is essentially self-explanatory and straightforward as it simply follows the algorithms and formulas in Section \ref{sec.2}. It consists of a single function {\tt{\small dead.alphas(hld.good, hld.dead, d, do.trunc = T)}}. Here {\tt{\small hld.dead}} is a 3-dimensional $N_{dead}\times M\times d$ array $P_{iAs}$ of stock positions for dead alphas labeled by $i \in J^\prime$ (see Section \ref{sec.2}); {\tt{\small d}} is the number of trading days over which the matrices are averaged in (\ref{X1}); {\tt{\small hld.good}} is a 2-dimensional $N_{good}\times M$ matrix $P^*_{iA} = P_{iAs}$ of stock positions for good alphas labeled by $i \in J$ for the most recent (``today's") date $s=1$ for which we wish to compute the desired holdings ${\widetilde P}_{iAs}$ (see Section \ref{sec.2}); {\tt{\small do.trunc}} controls whether the eRank is truncated or rounded. Internally the function {\tt{\small dead.alphas()}} calls the function {\tt{\small calc.erank()}}, which is given in Appendix A of \cite{Decode}. The output is an $(N_{good} + K) \times M$ matrix, whose first $N_{good}$ rows are the matrix ${\widetilde P}_{iAs}$  (for $s=1$), and the remaining rows are the $K$ principal components $V^{(a)}_A$, $a=1,\dots,K$, where $K$ is computed using eRank (see Section \ref{sec.2}). Here we assume that $M \ll N_{dead}$. If the situation is reversed, i.e, we have not too many dead alphas ($N_{dead} \ll M$), then it is more efficient to replace the built-in R function {\tt{\small eigen()}} (which is based on power iterations \cite{PowerIt}) by the function {\tt{\small qrm.calc.eigen.eff()}} given in Appendix C of \cite{StatRM}, which provides a much more efficient method than power iterations.\\
\\
{\tt{\small
\noindent dead.alphas <- function (hld.good, hld.dead, d, do.trunc = T)\\
\{\\
\indent x <- 0\\
\indent for(s in 1:d)\\
\indent \indent x <- x + t(hld.dead[, , s]) \%*\% hld.dead[, , s]\\
\\
\indent x <- eigen(x)\\
\indent x.val <- x\$values\\
\indent x.vec <- x\$vectors\\
\indent k <- calc.erank(x.val, excl.first = F)\\
\indent if(do.trunc)\\
\indent \indent k <- trunc(k)\\
\indent else\\
\indent \indent k <- round(k)\\
\\
\indent q <- x.vec[, 1:k]\\
\indent y <- t(hld.good)\\
\indent y <- residuals(lm(y $\sim$ -1 + q))\\
\indent y <- t(y) / colSums(abs(y))\\
\indent y <- rbind(y, t(q))\\
\indent return(y)\\
\}
}}

\section{DISCLAIMERS}\label{app.B}

{}Wherever the context so requires, the masculine gender includes the feminine and/or neuter, and the singular form includes the plural and {\em vice versa}. The author of this paper (``Author") and his affiliates including without limitation Quantigic$^\circledR$ Solutions LLC (``Author's Affiliates" or ``his Affiliates") make no implied or express warranties or any other representations whatsoever, including without limitation implied warranties of merchantability and fitness for a particular purpose, in connection with or with regard to the content of this paper including without limitation any code or algorithms contained herein (``Content").

{}The reader may use the Content solely at his/her/its own risk and the reader shall have no claims whatsoever against the Author or his Affiliates and the Author and his Affiliates shall have no liability whatsoever to the reader or any third party whatsoever for any loss, expense, opportunity cost, damages or any other adverse effects whatsoever relating to or arising from the use of the Content by the reader including without any limitation whatsoever: any direct, indirect, incidental, special, consequential or any other damages incurred by the reader, however caused and under any theory of liability; any loss of profit (whether incurred directly or indirectly), any loss of goodwill or reputation, any loss of data suffered, cost of procurement of substitute goods or services, or any other tangible or intangible loss; any reliance placed by the reader on the completeness, accuracy or existence of the Content or any other effect of using the Content; and any and all other adversities or negative effects the reader might encounter in using the Content irrespective of whether the Author or his Affiliates is or are or should have been aware of such adversities or negative effects.

{}The R code included in Appendix \ref{app.A} hereof is part of the copyrighted R code of Quantigic$^\circledR$ Solutions LLC and is provided herein with the express permission of Quantigic$^\circledR$ Solutions LLC. The copyright owner retains all rights, title and interest in and to its copyrighted source code included in Appendix \ref{app.A} hereof and any and all copyrights therefor.

\end{document}